\input harvmac
\newcount\figno
\figno=0
\def\fig#1#2#3{
\par\begingroup\parindent=0pt\leftskip=1cm\rightskip=1cm\parindent=0pt
\baselineskip=11pt
\global\advance\figno by 1
\midinsert
\epsfxsize=#3
\centerline{\epsfbox{#2}}
\vskip 12pt
{\bf Fig. \the\figno:} #1\par
\endinsert\endgroup\par
}
\def\figlabel#1{\xdef#1{\the\figno}}
\def\encadremath#1{\vbox{\hrule\hbox{\vrule\kern8pt\vbox{\kern8pt
\hbox{$\displaystyle #1$}\kern8pt}
\kern8pt\vrule}\hrule}}

\overfullrule=0pt

%

\def\R{{\bf R}}

\font\zfont = cmss10 

\def\bigone{\hbox{1\kern -.23em {\rm l}}}
\def\ZZ{\hbox{\zfont Z\kern-.4emZ}}

\Title{hep-th/9609159, IASSNS-HEP-96-97}
{\vbox{\centerline{PHYSICAL INTERPRETATION OF CERTAIN}
\bigskip
\centerline{ STRONG COUPLING SINGULARITIES}}}
\smallskip
\centerline{Edward Witten\foot{Research supported in part
by NSF  Grant  PHY-9513835.}}
\smallskip
\centerline{\it School of Natural Sciences, Institute for Advanced Study}
\centerline{\it Olden Lane, Princeton, NJ 08540, USA}\bigskip

\medskip

\noindent
We interpret certain strong coupling singularities of the 
$E_8\times E_8$ heterotic string on K3 in terms of
exotic six-dimensional theories in which $E_8$ is a gauge
symmetry.  These theories are closely related to theories
obtained at small instanton singularities, which have $E_8$ as
a global symmetry.
\Date{September, 1996}
\bigskip\noindent{\it Introduction}

By considering Type IIB superstring theory on K3, near an $A-D-E$
singularity, one obtains an exotic and previously unknown
six-dimensional theory \ref\witten{E. Witten, ``Some Comments On
String Dynamics,'' hepth/9507121.} from which gravity decouples.  
(The same theory can be constructed via parallel five-branes
\ref\strominger{A. Strominger, ``Open $p$-Branes,'' hepth/9512059.}.)
The limiting theory is a flat space, interacting,
super-Poincar\'e-invariant, theory which may well obey the axioms of 
local field theory but whose existence is certainly
not predicted by any standard 
field-theoretic construction.  (In fact, known constructions
above four dimensions appear to give only free field theories.)
One indication of the interest of this particular exotic theory
is that its existence has $S$-duality of four-dimensional $N=4$
super-Yang-Mills theory (for $A-D-E$ gauge groups) as a corollary
\witten.

\nref\hanany{O. Ganor and A. Hanany, ``Small $E_8$ Instantons And
Tensionless Noncritical Strings,'' hepth/9602120.}
\nref\sw{N. Seiberg and E. Witten, ``Comments On String Dynamics
In Six Dimensions,'' hepth/9603003. }
\nref\seiberg{N. Seiberg, ``Five-Dimensional SUSY Field Theories,
Non-trivial Fixed Points, and String Dynamics,'' hepth/9608111.}
\nref\sm{D. Morrison and N. Seiberg, ``Extremal Transitions
And Five-Dimensional Supersymmetric Field Theories,'' hepth/9609070.}
\nref\vetal{M. Douglas, S. Katz, and C. Vafa, ``Small Instantons,
Del Pezzo Surfaces, And Type $I'$ Theory,'' hepth/9609071.}
\nref\ganor{O. Ganor, ``Toroidal Compactification Of Heterotic
6D Non-Critical Strings Down To Four Dimensions,'' hepth/9608109.}
\nref\duff{M. J. Duff, R. Minasian, and E. Witten, ``Evidence
For Heterotic/Heterotic Duality,'' hepth/9601036.}
\nref\vm{D. Morrison and C. Vafa, ``Compactification of $F$-Theory
On Calabi-Yau Threefolds, I, II,'' hepth/9602114,9603161.}
\nref\phaset{E. WItten,
``Phase Transitions In $M$-Theory and $F$-Theory,'' hepth/9603150.}
Other similarly exotic theories may be constructed, for example,
by taking suitable limits of heterotic string compactification on K3.
By considering a small $E_8$ instanton, one gets a six-dimensional
theory \refs{\hanany,\sw}
in which $E_8$ appears as a {\it global} symmetry group.  This theory
has extremely rich properties, as shown in \refs{\sm - \ganor}.  By
increasing the heterotic string coupling to a critical value
\duff\ one obtains a somewhat analogous
critical point, most effectively studied using
$F$-theory \refs{\vm,\phaset}.  These theories have strings,
which can be constructed approximately  as solitons 
\ref\oduff{M. Duff, H. Lu, and
C. N. Pope, ``Heterotic Phase Transitions And Singularities Of
The Gauge Dyonic String,'' hepth/9603037,}, whose tension vanishes
at the critical point.

In fact, the K3 constructions each give several theories.  By considering
$n$ nearly coincident small $E_8$ instantons, one gets a theory
that we will call $V_n$.  For reasons that we will review in section
two, $V_n$ is a theory with $E_8$ as a {\it global} symmetry group.
To obtain exotic theories with {\it local} gauge symmetry in
six dimensions, we can turn to the strong coupling singularities.
Consider the $E_8\times E_8$ heterotic string on K3, with instanton
numbers $(24-n,n)$ in the two factors.  If $n=12$ there is
no strong coupling singularity \duff; for $n\not=12$ there is
(upon possibly exchanging the two factors) no loss of generality in
assuming that $n<12$.  The $(24-n,n)$ theory with $n<12$ has
for its strong coupling singularity a theory that we will call
$T_n$.

We will argue that the $T_n$'s should be interpreted as theories
with {\it local} $E_8$ gauge symmetry.  Among these,
 we will interrpret $T_0$ as  a sort of (almost) ``pure''
gauge theory with ``no charged matter fields,'' Given such a theory,
one can try to couple to ``matter,'' that is to a theory with
$E_8$ as a global symmetry.  In six dimensions, the only theories
we know with $E_8$ global symmetry are the $V_n$'s; we will argue
that at least heuristically, the theories     $T_n$
for $1\leq n\leq 11$ can be interpreted as a ``gauging'' of
$V_n$ by coupling to the (almost) ``pure gauge theory'' $T_0$. 

\bigskip\noindent
{\it Review Of Small Instanton Theories}

First, we reconsider the theory that arises from a small $E_8$ instanton
in heterotic string compactification on $\R^6\times {\rm K3}$.

\def\K3{{\rm K3}}
A small instanton near a point
 $p\in \K3$ can be interpreted as a five-brane whose world-volume
fills out the six-manifold $Q=\R^6\times\{p\}$.  Certain modes propagate
on $Q$.  To claim that the theory on $Q$ is a {\it flat space theory}
with {\it $E_8$ as a global symmetry}, the key point is to show that both
the gravitons and the $E_8$ gauge bosons can be made to decouple from the
theory on $Q$, to any desired precision.  This may be done simply by
increasing the volume of the K3.  The graviton and gauge boson wave
functions then spread all over K3, and as the K3 grows, they decouple
from the theory on $Q$.    The fact that the $E_8$ gauge bosons decouple
from the theory on $Q$ means that in studying this theory, $E_8$ can
be treated as a global symmetry.  The fact that the graviton decouples
means that this is likewise a theory in which the Poincar\'e group is
a global symmetry and there is no dynamical gravity, 
something which we have loosely described by calling
the theory a flat space theory.

The small instanton theory, which we will call $V_1$,
has a Higgs branch with 29 massless
hypermultiplets (representing the size and $E_8$ 
orientation of the instanton), 
and a Coulomb branch with one massless tensor multiplet (corresponding
to a situation in which the instanton has been converted to an $M$-theory
five-brane).
One could add a thirtieth massless hypermultiplet, representing the
instanton center of mass, but this would be a decoupled free field
and we choose not to include it.

By considering $n$ nearly coincident small $E_8$ instantons, one can
construct further theories that we will call $V_n$.  $V_n$ has a 
Higgs branch with $30n-1$ massless hypermultiplets (which parametrize
the sizes, $E_8$ orientations, and relative positions of the instantons), 
a Coulomb branch
with $n$ massless tensor multiplets (the branch with $n$ $M$-theory 
five-branes), and various mixed branches with some of each.  Notice
that on the Higgs branch, 
if we consider a situation in which the instantons
are far apart compared to their sizes, then the theory $V_n$ reduces
to $V_1^n$ ($n$ copies of $V_1$) plus $n-1$ free massless hypermultiplets
(the relative positions of the instantons).  Other analogous
reductions arise in limits in which some but not all of the separations
between instantons are large compared to the instanton sizes.

\bigskip
\noindent
{\it Search For Gauge Symmetry}

Now if we want to find an analogous theory with $E_8$ as a {\it gauge
symmetry}, we must find a singularity at which the $E_8$ gauge bosons
do {\it not} 
decouple.  One option is the strong coupling singularity
of the $E_8\times E_8$ heterotic string with instanton numbers $(24-n,n)$,
$n\leq 12$.  Since it is hard to understand this singularity from the
heterotic string point of view, it is not obvious from this point of
view whether or not
the gauge bosons are decoupled from whatever theory may emerge at the
singularity.  It is likewise not clear whether gravity is so decoupled.

What settles this 
question is the approach via $F$-theory \vm.  The $(24-n,n)$
heterotic string is equivalent to $F$-theory on the Hirzebruch surface
${\bf F}_{12-n}$.  In this description, the strong coupling singularity
is localized on a 
curve $C\subset {\bf F}_{12-n}$.  ($C$ is the exceptional
section of the Hirzebruch surface, with $C\cdot C=-12+n$.)  Since the
graviton wave functions 
are meanwhile spread over the whole Hirzebruch surface,
it follows that gravity is decoupled from the theory that emerges at the
singularity, so 
that this theory (if it is a field theory at all) is a flat
space quantum field theory.  On the other hand, according to \vm, the
gauge bosons of one
of the two $E_8$'s (namely the one in which the instanton number is $n$)
are localized on the same curve $C$, leading us to expect that
they are not decoupled from whatever physics is supported on $C$.
Therefore, the theory -- which we will call $T_n$ -- associated
with the strong coupling singularity of the $(24-n,n)$ heterotic string
is potentially an $E_8$ gauge theory.  

One may wonder whether in some situation gravity would fail to decouple
from the modes that become strongly coupled at such a singularity,
giving some sort of exotic phase of quantum general relativity with
novel long distance behavior.  Whether this is possible is not clear. 

\bigskip\noindent{\it Identifying The ``Matter Content''}

Going back to the $T_n$'s, we would like to say more about their
``matter content.''  What is the unbroken gauge group in the theory $T_n$?
This depends on where we are in moduli space.  For any given value
of $n$, in the $(24-n,n)$ heterotic string compactification
on K3, 
there is a generic unbroken subgroup of $E_8$ (that is, of the
``second'' $E_8$ with instanton number $n$). We will call this generic
unbroken group   
$G_n$.  For instance (on the known branches of the moduli spaces),
$G_0=G_1=G_2=G_3=E_8$, 
$G_4=E_7$, $G_8=SO(8)$, $G_9=SU(3)$, and $G_{10}$ and $G_{11}$
are trivial.  (We stop the list at $n=11$, of course, since that is the
last example for the strong coupling singularity that we are studying here.)
On the other hand, at special loci in moduli space, it is possible to restore
part or all of the $E_8$.

We interpret this to mean that in some sense the $T_n$ theories are
$E_8$ gauge theories coupled to matter.  We propose that the $T_0$ theory,
in which the $E_8$ is always unbroken, is roughly comparable to a ``pure
gauge theory.''\foot{The $T_1$, $T_2$, and $T_3$ theories also
have always unbroken $E_8$, but will be given a slightly different
interpretation momentarily.}  This statement requires
some explanation.   In addition to the $E_8$ vector multiplets, the $T_0$
theory has a massless tensor multiplet (whose scalar parametrizes the
heterotic string coupling constant) with a moduli space $\R^+$.  The
critical point is the endpoint of the $\R^+$, and at that point, since
the physics is not free in the infrared, the particle spectrum and gauge
group are likely to be ill-defined.  However, at a generic point on $\R^+$,
away from the endpoint, the physics is infrared-free, the spectrum makes
sense, and the gauge group is well-defined and equal to $E_8$.  
In calling the theory an (almost) ``pure gauge theory,'' we mean simply
that away from the critical point,
there is no charged matter; the generic
massless spectrum consists of the $E_8$
vector multiplet plus one tensor multiplet.

Now, if the $T_0$ theory is thought of as an $E_8$
gauge theory, one might
expect to be able to couple it to a ``matter system,'' that is a field
theory with $E_8$ global symmetry, which one might try to gauge.  
One set of theories with global $E_8$ symmetry are the small instanton
theories $V_n$, $n\geq 1$.  We propose that the $T_n$ theories
for $n=1,\dots,11$ are obtained by ``gauging'' the $V_n$, that
is by coupling them to the ``pure gauge theory'' $T_0$.  
(The fact that the $T_n$ series apparently ends at $n=11$ would from
this point of view mean that for $n>11$ one has ``too much matter''
to obtain a well-defined theory; the bound on the amount of matter
would be analogous to the familiar asymptotic freedom bound in four
dimensions, or the less familiar bound found in five dimensions in
\seiberg.) 

As a check on this, we may try to show that the theory obtained
by ``gauging'' $V_n$ would have the same generic unbroken gauge
group on the Higgs branch as the $(24-n,n)$ heterotic string.
In fact, this is almost a tautology.  If we gauge $V_n$, the moduli
space of the Higgs branch is obtained by starting with  the moduli space
${\cal M}_n$ of $n$ $E_8$ instantons on $\R^4$, and then taking
the ``hyper-Kahler quotient'' of ${\cal M}_n$ by $E_8$.  (The hyper-Kahler
quotient is obtained by setting the $D$-fields to zero and then dividing
by $E_8$.  $D$-fields, sometimes called the components of the
hyper-Kahler moment map, are defined for any
hyper-Kahler manifold -- in this case the Higgs branch of $V_n$ --
with a simple group action -- in this case $E_8$.
One must set the $D$-fields to zero to find a supersymmetric
vacuum because the potential energy of the supersymmetric gauge theory
is a multiple of $\vec D^2$.)   
On the other hand, if we consider $n$ almost coincident
small $E_8$ instantons on K3, the moduli space of such objects is
precisely the same as the hyper-Kahler quotient of ${\cal M}_n$ by $E_8$
\foot{
Since K3 is compact, one automatically divides by $E_8$ in constructing
instanton moduli space on K3.  Supersymmetry relates the operation of
dividing by $E_8$ to that of setting to zero the $D$ fields, which
therefore necessarily is implicit in the moduli problem of instantons
on K3.  It is possible by standards arguments
to show explicitly how vanishing of the $D$ fields
comes in in the small instanton limit, but this will not be done here.
} (assuming that one suppresses the overall center of mass position of the
instantons, which is parametrized by $\R^4$ in one case and K3 in the other).
Therefore, not only does the theory obtained heuristically by ``gauging''
$V_n$ have for its generic unbroken gauge group on the Higgs branch
the group $G_n$ that appears in the $(24-n,n)$ heterotic string;
the Higgs branches actually coincide.

The fact that the theories $T_i$ for $i=1,2,3$ have the same unbroken
gauge group as $T_0$ is now clear intuitively: they have massless
charged matter fields but not enough of them to generate a Higgs branch.

Now let us ask what is the smallest value of $n$ for which complete
Higgsing of $E_8$ is possible in the $T_n$ theory.  From what has been
said, this is the same as the smallest value of $n$ for which complete
Higgsing is possible in the $(24-n,n)$ heterotic string, namely $n=10$.

Now consider the theory $T_{11}$.  We interpret this as the ``gauging''
of the theory $V_{11}$ of 11 small instantons.  In a suitable limit
(with one small instanton far from the other ten) $V_{11}$ reduces
to $V_1\times V_{10}$ (with an additional and irrelevant massless free
hypermultiplet that parametrizes the relative separation of the ten instantons
from the eleventh one).  Therefore, in this limit, $T_{11}$ reduces to the
``gauging'' of $V_1\times V_{10}$.  The ``gauging'' of $V_{10}$ gives
a theory in which complete Higgsing is possible.  Upon going to a region of
$V_{10}$ moduli space in which this complete Higgsing occurs, the $E_8$ gauge
bosons all get mass.  The low energy theory in this region consists
of some infrared-free massless hypermultiplets plus a copy of the
 theory $V_1$, which near its critical point is an interacting theory.

We have deduced, therefore, that the theory $T_{11}$, on a suitable
part of its Higgs branch, reduces to the theory $V_1$.
This corresponds to a familiar fact \witten: the strong coupling
singularity of the $(13,11)$ heterotic string is actually governed
by the small instanton theory $V_1$.

This then is perhaps the one fact that we have explained in this paper:
the value of $n$ at which the $(24-n,n)$ heterotic string has $V_1$
for its strong coupling singularity (namely $n=11$) is one more than
the smallest value (namely $n=10$) at which complete Higgsing is possible.

\listrefs
\end